\documentclass[aps,prb,twocolumn,groupedaddress,showpacs]{revtex4-1}
\usepackage[T1]{fontenc}
\usepackage{textcomp}
\usepackage{lmodern}
\usepackage[latin1]{inputenc}
\usepackage[swedish,english]{babel}
\usepackage{sidecap}
\usepackage{amsmath}
\usepackage{subfig}
\usepackage{amsfonts}
\usepackage{graphicx} 
\usepackage{multirow}
\usepackage{soul}
\usepackage[retainorgcmds]{IEEEtrantools}

\begin{document}

\title{Three and Four-Body Intervortex Forces in the Ginzburg-Landau Models of Single- and Multicomponent Superconductivity}

\author{Alexander Edstr\"om}
\affiliation{KTH}

\date{\today}

\begin{abstract}
A systematic numerical study of non-pairwise vortex interaction forces in the Ginzburg-Landau model for single- and multicomponent  superconductivity is presented. The interactions are obtained by highly accurate  numerical free energy minimization. In particular a three-body interaction is defined as the difference between the total interaction and sum of pairwise interactions in a system of three vortices and such interactions are studied for single and two-component type-1, type-2, and  type-1.5 superconductors. In the investigated regimes, the three-body interaction is found to be short-range repulsive but long-range attractive in the type-1 case, zero in the critical $\kappa$ (Bogomoln'y) case, attractive in the type-2 case and repulsive in the type-1.5 case. Some systems of four vortices are also studied and results indicate that four-body forces are of substantially less significance than the three-body interactions. 
\end{abstract}


\maketitle

\section{Introduction}

Since the discovery by Onsager in superfluids and by Abrikosov in superconductors \cite{Abrikosov}, vortices have been found to play crucial role in many aspects of superconductivity and superfluidity \cite{Tinkham}. The main difference which can be observed in different types of superconductors is how the vortices interact and the interaction between a vortex pair is known to be attractive for type-1 but repulsive for type-2 superconductors \cite{PhysRevB.3.3821}. In a multicomponent superconducting system it is possible to have also type-1.5 behaviour with short-range repulsive but long-range attractive interactions \cite{PhysRevB.72.180502,PhysRevLett.102.117001,PhysRevB.84.134515,PhysRevB.83.174509,Babaev20122,PhysRevB.83.020503,PhysRevB.81.020506,PhysRevB.85.094511,2010PhRvL.105f7003B,PhysRevB.86.060513,2011PhRvB..84i4515S,2012PhRvB..85m4514S,PhysRevB.84.134518,2011PhRvB..83v4522L,2010PhRvB..81u4514G}, for a review see \cite{Babaev20122}. While type-2 superconductors exhibit a vortex state with vortices in an Abrikosov lattice, type-1.5 superconductors also exhibit a semi-meissner state where the non-monotonic interactions lead to more complex formations such as vortex clusters. 

Non-linearity in the Ginzburg-Landau model implies that interactions between many vortices cannot be described simply as the sum of pairwise interactions, although it is a good approximation for forces between widely separated vortices. It can lead to quantitative changes in vortex lattices in type-2 superconductors. It was suggested in \cite{PhysRevB.84.134515} that non-pairwise contributions to interactions between many vortices in certain type-1.5 systems can dramatically affect vortex configurations. The possibility of non-trivial effects caused by non-pairwise interactions motivates a further study of such interactions as done here. In particular the three-body interaction defined as the difference between total and sum of pairwise interactions between three vortices will be studied. Also in the more simple single-component type-1 and type-2 cases, the form of three-body interactions are known only for a very limited set of configurations. 
In \cite{PhysRevB.83.054516} the problem of three vortices in a single-component type-2 and type-1 system were studied only for the case of  an equilateral triangle. In the two-component case, three-body intervortex forces were studied for one fixed position of two vortices as a function of arbitrary position of a third vortex only \cite{PhysRevB.84.134515}. Thus multibody intervortex forces are known only for few very limited set of configurations and it warrants a more extensive study. It will therefore be investigated in Sec. \ref{sec:type1}, Sec. \ref{sec:type2} and Sec. \ref{sec:type15} for a wider set of configurations than previously studied. We also present the first study of four-body forces thereby assessing the scaling of multibody forces when extra vortices are included. The known analytical suppression of three-body interactions in the Bogomoln'y limit \cite{Manton}, $\kappa = \frac{1}{\sqrt{2}}$,  will be used as a benchmark to check the accuracy of our numerical scheme in Sec. \ref{sec:critkap} and systems of four vortices are studied in Sec. \ref{sec:4bint}.

The free energy density of a two-component superconductor in Ginzburg-Landau theory is 
\begin{IEEEeqnarray}{rCl}
	f & = & \frac{1}{2} \sum_{j=1,2} \left[ \left| \left( \nabla + i q {\mathbf A} \right) \psi_{j} \right|^{2} + \left( 2\alpha_{j} + \beta_j |\psi_j|^2  \right) |\psi_j|^2 \right] + \nonumber
\\ &&  + \frac{1}{2}\left( \nabla \times {\mathbf A} \right)^2 - \eta |\psi_1||\psi_2|\cos(\varphi_2 - \varphi_1) ,
	\label{eq:f}
\end{IEEEeqnarray} 
where q is the electric charge of the superconducting charge carriers, $\psi_j = |\psi_j|e^{i\varphi_j}$ are the order parameters of the two components, ${\mathbf A}$ is the magnetic vector potential and $\eta$ is the Josephson coupling. $\alpha_j$ and $\beta_j$ are Ginzburg-Landau coefficients corresponding to each component. To determine intervortex forces the free energy in Eq. \ref{eq:f} should be minimized with respect to both of the order parameters as well as the vector potential. This is done here numerically using a finite difference method developed in \cite{PhysRevB.84.134515} and briefly described in Sec. \ref{sec:num}. Single-component systems are described by the same free energy except without the terms corresponding to the second condensate and the coupling. Physical realization of two-component models like the one above are multiband superconductors \cite{0370-1328-84-4-313,2010PhRvB..81u4514G,2012PhRvB..85m4514S,PhysRevLett.89.067001}
and projected superconducting states of metallic hydrogen \cite{j.nuclphysb.2004.02.021,nature}.

\section{Numerical Method}\label{sec:num}

The numerical method used to calculate intervortex forces is essentially that used also in \cite{PhysRevB.84.134515}. A starting guess is done for ${\mathbf A}$ and $\psi_j$, setting the phase winding and positions of vortices. The free energy in Eq. \ref{eq:f} is then minimized by Newton-Raphson based iteration. Vortex pinning is implemented by setting $\psi = 0$ at the position of vortices which is minimally invasive, and thus highly accurate for determining multibody forces, but does not work for too short vortex separations as vortices will then tend to escape from their positions while leaving singular points with $\psi_j = 0$ at the original positions due to the pinning constraint. Whether this happens can easily be checked by plotting the solutions for ${\mathbf A}$ and $\psi_j$.  

The energy was minimized for a rectangular grid with size $N_1$ until it converged to a value $E(N_1)$. This was then repeated for a new grid size $N_2$, typically a factor four (double in x- and y-direction) greater,  to give an energy $E(N_2)$ etc. This allowed checking for convergence with respect to grid size. Final grid sizes were in the order $N_n \sim 10^7$, giving a relative convergence 
\begin{equation}
	\frac{E(N_n) - E(N_{n-1})}{E(N_n)} < 10^{-5}. 
\end{equation}

Interaction energies are calculated by first finding the energy $E_1$ of a system with only a single vortex. The pairwise interaction energy $E(R)$ of a vortex pair with intervortex distance $R$ is calculated by finding the total energy $E_{\text{tot}}(R)$ of the system with two vortices and then subtracting the single vortex energies so 
\begin{equation}
	E(R) = E_{\text{tot}}(R) - 2 E_1. 
\end{equation}
Similarly, if the total energy in a system of three vortices with intervortex distances $R_1$, $R_2$ and $R_3$ is $E_{\text{tot}}(R_1,R_2,R_3)$, the total interaction energy of the system is 
\begin{equation}
	E_{\text{int}}(R_1,R_2,R_3)  = E_{\text{tot}}(R_1,R_2,R_3) - 3 E_1
\end{equation}
and the three-body interaction is the difference between total interaction and sum of pairwise interactions
\begin{equation}
	E_{3}(R_1,R_2,R_3)  = E_{\text{int}}(R_1,R_2,R_3) - E(R_1) - E(R_2) - E(R_3). 
\end{equation}

\section{Pairwise Interactions}

For the asymptotic long-range pairwise interaction between vortices in a single-band system, analytical results are available \cite{PhysRevB.3.3821,PhysRevD.55.3830}. Introducing $\psi = \sqrt{\frac{|\alpha|}{\beta}}f \text{e}^{\text{i} \varphi}$, ${\mathbf Q} = {\mathbf A} - \frac{\nabla \varphi}{\kappa}$ and the Ginzburg-Landau parameter $\kappa = \sqrt{\frac{2\beta}{q^2}}$, the free energy is
\begin{equation}
	F = \int \text{d}V \left[ \frac{1}{2}\left(1-f^2\right)^2 + \left( \frac{\nabla f}{\kappa} \right)^2 + {\mathbf Q}^2f^2 + \left( \nabla \times {\mathbf Q}\right) \right].
\label{eq:f2}
\end{equation}
Minimization of the energy leads to the Ginzburg-Landau equations 
\begin{equation}
	 \left[ {\mathbf Q}^2 - \left( \frac{\nabla}{\kappa}\right)^2\right] f = f\left( 1 - f^2\right)
\label{eq:GL1}
\end{equation}
and
\begin{equation}
	 \nabla \times \nabla \times {\mathbf Q} + f^2 {\mathbf Q} = 0.
\label{eq:GL2}
\end{equation}
The cylindrically symmetric solutions $f_0$ and ${\mathbf Q}_0$ to Eq. \ref{eq:GL1} and Eq. \ref{eq:GL2} for a single vortex with boundary constraints 
\begin{equation}
	 f_0 \rightarrow 0 \quad \text{and} \quad {\mathbf Q}_0 r \rightarrow \kappa \quad \text{as} \quad r \rightarrow 0
\label{eq:boundshort}
\end{equation}
\begin{equation}
	 f_0 \rightarrow 1 \quad \text{and} \quad {\mathbf Q}_0 r \rightarrow 0 \quad \text{as} \quad r \rightarrow \infty 
\label{eq:boundlong}
\end{equation}
are for large distances $r$
\begin{equation}
	 {\mathbf Q}_0 = c K_1 (r) \hat{\theta} 
\label{eq:Q0}
\end{equation}
and
\begin{equation}
	 g = 1 - f_0 = d K_0 (\sqrt{2} \kappa r),
\label{eq:f0}
\end{equation}
where $K_0$ and $K_1$ are modified Bessel functions while $c$ and $d$ are some values dependent on $\kappa$.

In the presence of a second vortex, far away from the first one, the solutions to Eq. \ref{eq:GL1} and Eq. \ref{eq:GL2} change to 
\begin{equation}
	 {\mathbf Q} = {\mathbf Q}_0 + {\mathbf Q}_1,
\label{eq:Q}
\end{equation}
\begin{equation}
	 f = f_0 + f_1,
\label{eq:f0f1}
\end{equation}
where ${\mathbf Q}_1$ and $f_1$ are small perturbations. Inserting Eq. \ref{eq:Q} and Eq. \ref{eq:f0f1} into Eq. \ref{eq:GL1} and Eq. \ref{eq:GL2} yields perturbation equations for ${\mathbf Q}_1$ and $f_1$ and upon linearization the solutions are 
\begin{equation}
	 {\mathbf Q}_1 (\hat{{\mathbf r}})= {\mathbf Q}_0 ({\mathbf r} - {\mathbf r}'),
\label{eq:Q1}
\end{equation}
\begin{equation}
	 f_1 ({\mathbf r}) = g({\mathbf r} - {\mathbf r}'),
\label{eq:f1}
\end{equation}
where ${\mathbf r}'$ is the position of the second vortex. That is, the cylindrically symmetric solution for the first vortex is perturbed by the asymptotic long-range solution for the second vortex. 

To find the interaction energy, Eq. \ref{eq:Q} and Eq. \ref{eq:f0f1} are inserted into Eq. \ref{eq:f2}. This yields 
\begin{equation}
	F = U_0 + U_\text{int},
\label{eq:ftot}
\end{equation}
where $U_0$ is the single vortex energy 
\begin{equation}
	U_0 = \int \text{d}V \left[ \frac{1}{2}\left(1-f_0^2\right)^2 + \left( \frac{\nabla f_0}{\kappa} \right)^2 + {\mathbf Q}_0^2f_0^2 + \left( \nabla \times {\mathbf Q}_0\right) \right]
\label{eq:U0}
\end{equation}
and $U_\text{int}$ is the interaction energy which consists of two terms so $U_\text{int} = U_1 + U_2$ where 
\begin{equation}
	U_1 = 2 \oint \text{d}{\mathbf S} \cdot \left[ {\mathbf Q}_1 \times \left( \nabla \times {\mathbf Q}_0 \right) \right] 
\label{eq:U1int}
\end{equation}
\begin{equation}
	U_2 = 2 \oint \text{d}{\mathbf S} \cdot \left( f_1 \frac{\nabla f_0}{\kappa} \right)
\label{eq:U2int}
\end{equation}
and integration is done over a surface containing the first vortex and far away from both vortices. Evaluation of these integrals yields
\begin{equation}
	U_1 (r)= 2\pi c^2 K_0 (r)
\label{eq:U1}
\end{equation}
\begin{equation}
	U_2 (r)= -\frac{2 \pi d^2}{\kappa} K_0 (\sqrt{2} \kappa r),
\label{eq:U2}
\end{equation}
where $r$ is the intervortex distance. Here $U_1$ is a repulsive electromagnetic interaction and $U_2$ is an attraction due to overlapping of the vortex cores.

It can be shown \cite{PhysRevB.3.3821}  that 
\begin{align}
	& d < \frac{c}{\sqrt{2}} \quad \text{for} \quad \kappa > \frac{1}{\sqrt{2}}, \nonumber \\ 
	& d = \frac{c}{\sqrt{2}} \quad \text{for} \quad \kappa = \frac{1}{\sqrt{2}}, \nonumber \\
	& d > \frac{c}{\sqrt{2}} \quad \text{for} \quad \kappa < \frac{1}{\sqrt{2}}.
\end{align}
Hence there is an attraction between type-1 vortices ($\kappa < \frac{1}{\sqrt{2}}$), repulsion between type-2 vortices ($\kappa > \frac{1}{\sqrt{2}}$) and no interaction between vortices in the Bogomoln'y limit ($\kappa = \frac{1}{\sqrt{2}}$).

\section{Results}

This section contains results of the numerical calculations of three and four-body interactions according to the method in Sec. \ref{sec:num}. Three-body interactions are presented for single-component type-1, critical $\kappa$ and type-2 as well as two-component type-1.5 systems in Sec. \ref{sec:type1} - \ref{sec:type15} respectively and four-body interactions are presented in Sec. \ref{sec:4bint}. Three-body interaction energies are presented in Fig. \ref{type1tb}, Fig. \ref{type2tb} and Fig. \ref{type15tb} as function of the position of the third vortex when a first vortex pair is fixed at positions $(\pm \frac{R_1}{2},0)$. Four-body interactions are shown as function of side length $R$ in a square configuration in Sec. \ref{sec:4bint}. All energies are given in terms of the energy of a single vortex in the system studied.  The "minimally invasive" vortex pinning mechanism does, as explained in Sec. \ref{sec:num}, not work for short distances and hence no data is available for short vortex separations, which is the reason for the empty regions in Fig. \ref{type1tb} - Fig. \ref{type15tbC}. 

\subsection{Single-Component Type-1 Three-Body Interaction}\label{sec:type1}

\begin{figure*}[!hbtp]
	\subfloat[][Total interaction.]{\label{totR1t1}\includegraphics[width=0.34\textwidth]{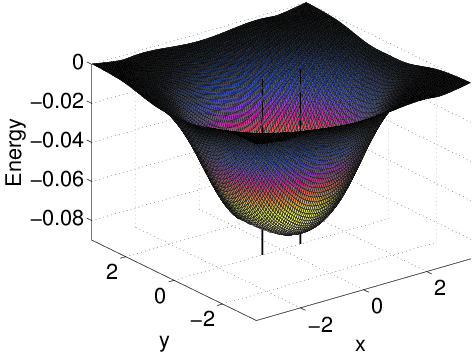}}
	\subfloat[][Total interaction.]{\label{totR2t1}\includegraphics[width=0.34\textwidth]{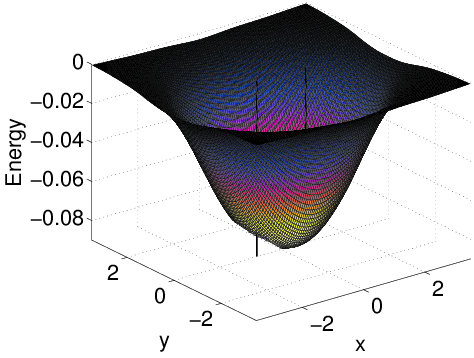}}
	\subfloat[][Total interaction.]{\label{totR3t1}\includegraphics[width=0.34\textwidth]{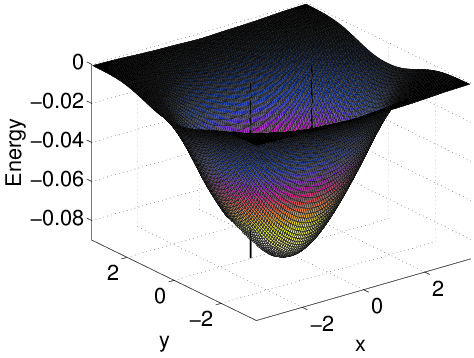}}\linebreak
	\subfloat[][Sum of pairwise interactions.]{\label{pairR1t1}\includegraphics[width=0.34\textwidth]{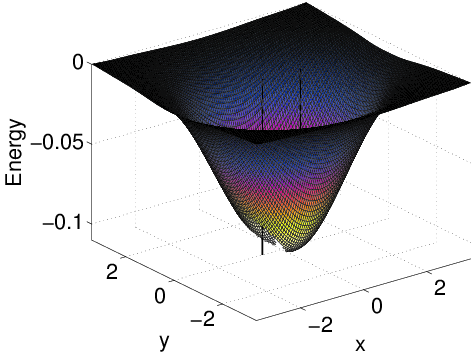}}
	\subfloat[][Sum of pairwise interactions.]{\label{pairR2t1}\includegraphics[width=0.34\textwidth]{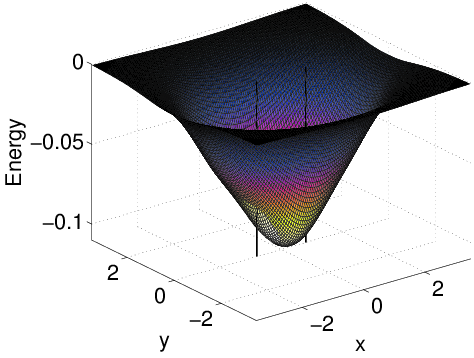}}
	\subfloat[][Sum of pairwise interactions.]{\label{pairR3t1}\includegraphics[width=0.34\textwidth]{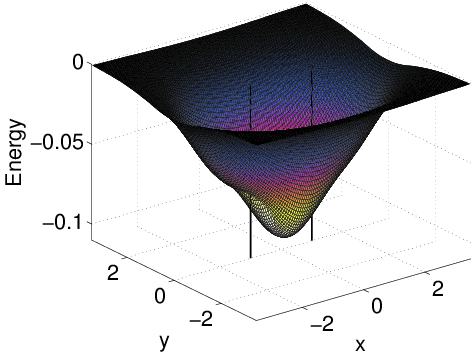}}\linebreak
	\subfloat[][Three-body interaction. $R_{1}=1.2$.]{\label{3bR1t1}\includegraphics[width=0.34\textwidth]{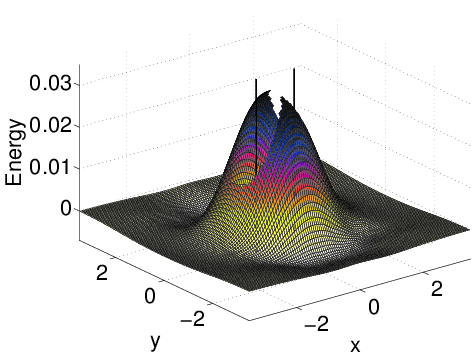}}
	\subfloat[][Three-body interaction. $R_{1}=1.6$.]{\label{3bR2t1}\includegraphics[width=0.34\textwidth]{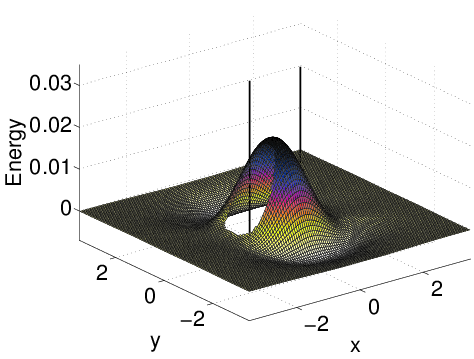}}
	\subfloat[][Three-body interaction. $R_{1}=2.0$.]{\label{3bR3t1}\includegraphics[width=0.34\textwidth]{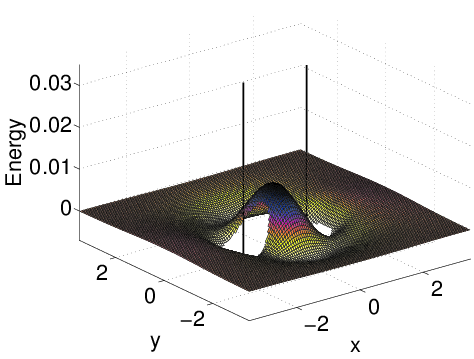}}
	\caption{Interactions in a single-component type-1 superconductor with Ginzburg-Landau parameters $\alpha = -1$, $\beta=1$ and $q=2.5$. Energy is shown as function of position of the third vortex with a first vortex pair fixed at $(\pm\frac{R_1}{2},0)$ (shown by black lines). Each column corresponds to a certain value of $R_1$.}
	\label{type1tb}
\end{figure*}

\begin{figure*}[!hbtp]
	\subfloat[][Total interaction. ]{\label{3bR1t1C}\includegraphics[width=0.34\textwidth]{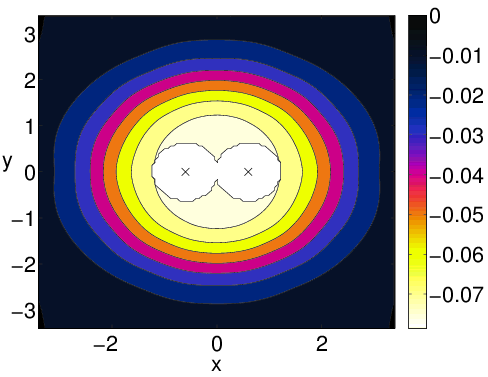}}
	\subfloat[][Total interaction. ]{\label{3bR1t1C}\includegraphics[width=0.34\textwidth]{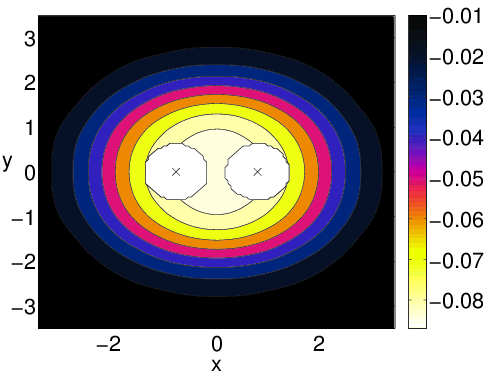}}
	\subfloat[][Total interaction. ]{\label{3bR1t1C}\includegraphics[width=0.34\textwidth]{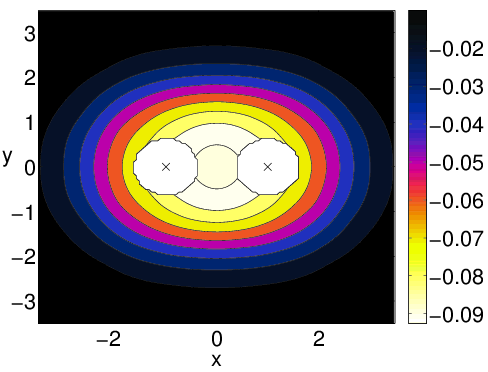}}\linebreak
	\subfloat[][Three-body interaction. $R_{1}=1.2$.]{\label{3bR1t1C}\includegraphics[width=0.34\textwidth]{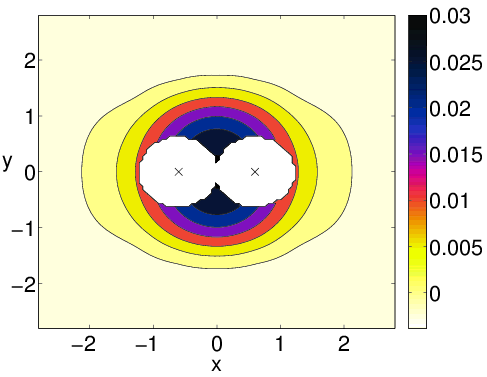}}
	\subfloat[][Three-body interaction. $R_{1}=1.6$.]{\label{3bR2t1C}\includegraphics[width=0.34\textwidth]{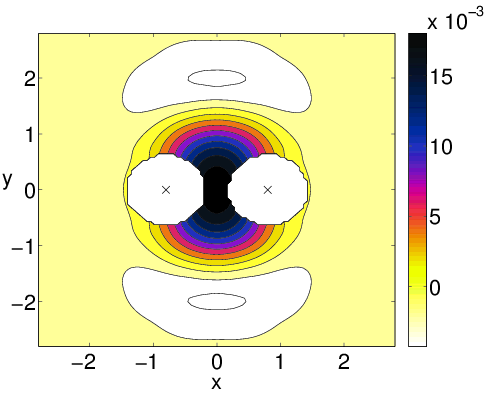}}
	\subfloat[][Three-body interaction. $R_{1}=2.0$.]{\label{3bR3t1C}\includegraphics[width=0.34\textwidth]{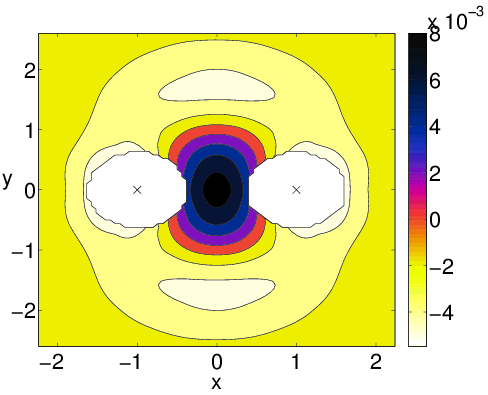}}
	\caption{Contour plots of the data from Fig. \ref{type1tb}.}
	\label{type1tbC}
\end{figure*}

Fig. \ref{type1tb} shows the total interaction energy, the sum of pairwise interaction energies and the three-body interaction energy of a single-component type-1 system. The third row is hence the difference between the first two rows. The parameters used are $\alpha = -1$, $\beta=1$ and $q=2.5$. Fig. \ref{type1tbC} shows the same data for total interaction and three-body interaction as that in Fig. \ref{type1tb} but as contour plots.

What can be noted from the results in Fig. \ref{type1tb} is that the three-body interaction is repulsive when all vortices are close together but also has an attractive region at longer vortex separations. The pairwise interaction is however attractive at any distance as expected from known analytical results. The plots for different values of $R_1$ show how the maximum magnitude of the three-body interaction decreases with increasing $R_1$. The magnitude of the repulsive part in the three-body interaction appears to decrease more strongly with increasing $R_1$ than the attractive region does. 

\subsection{Single-Component Critical Kappa Three-Body Interaction}\label{sec:critkap}

It can be shown analytically that the pairwise interaction energy is zero between a vortex pair in a superconductor with $\kappa$ at the critical value $\kappa_{\text{c}} = \frac{1}{\sqrt{2}}$ \cite{PhysRevB.3.3821,bogomolny}. This is consistent with numerical results obtained here by calculations of interaction energies in a critical $\kappa$ superconductor. Also \cite{Manton} there is no multibody interaction between vortices in a system with critical $\kappa$. Numerical results obtained here suggest that indeed no such three-body interaction exists in a critical $\kappa$ system, at least within the order of $10^{-5}E_1$ where $E_1$ is the energy of a single vortex. Thus this gives an estimate of the accuracy of our numerical scheme. The case studied is a single-component system with parameters $\alpha = -1$, $\beta = 1$ and $q = 2$.

\subsection{Single-Component Type-2 Three-Body Interaction}\label{sec:type2}

\begin{figure*}[!hbtp]
	\subfloat[][Total interaction.]{\label{totR1t1}\includegraphics[width=0.34\textwidth]{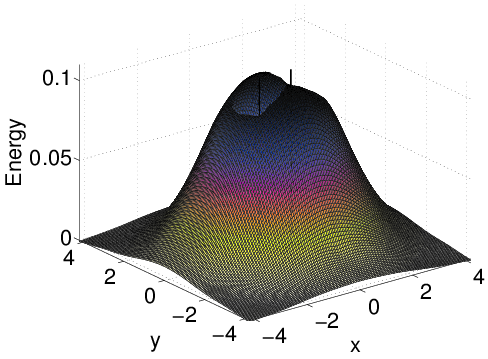}}
	\subfloat[][Total interaction.]{\label{totR2t1}\includegraphics[width=0.34\textwidth]{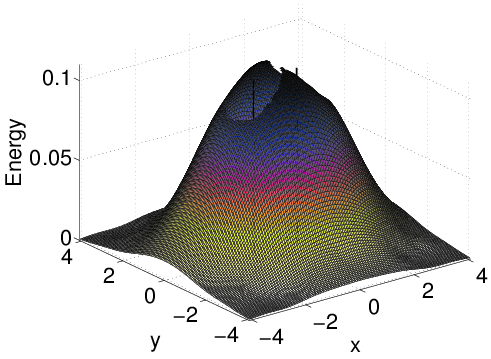}}
	\subfloat[][Total interaction.]{\label{totR3t1}\includegraphics[width=0.34\textwidth]{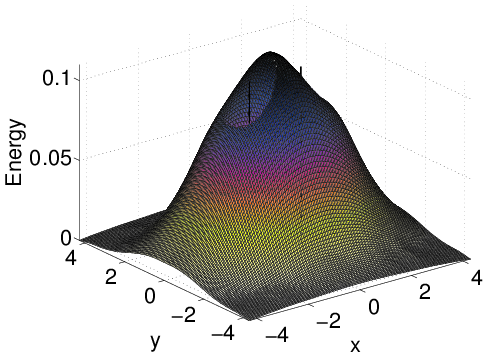}}\linebreak
	\subfloat[][Sum of pairwise interactions.]{\label{pairR1t1}\includegraphics[width=0.34\textwidth]{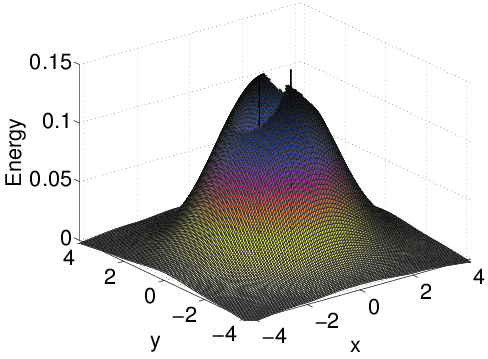}}
	\subfloat[][Sum of pairwise interactions.]{\label{pairR2t1}\includegraphics[width=0.34\textwidth]{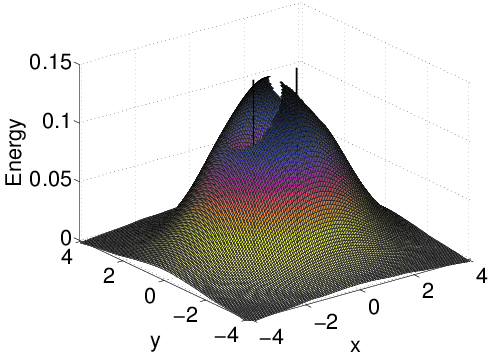}}
	\subfloat[][Sum of pairwise interactions.]{\label{pairR3t1}\includegraphics[width=0.34\textwidth]{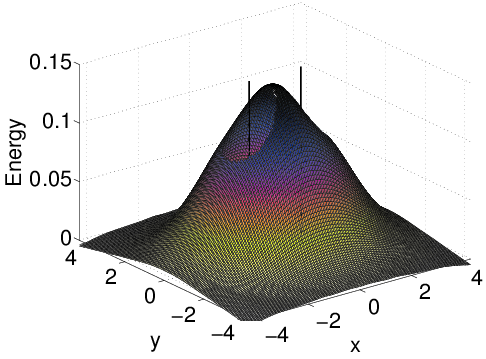}}\linebreak
	\subfloat[][Three-body interaction. $R_{1}=1.2$]{\label{3bR1t1}\includegraphics[width=0.34\textwidth]{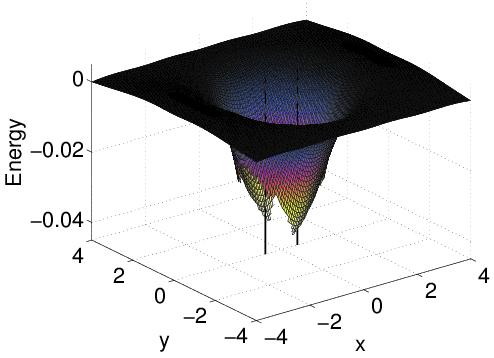}}
	\subfloat[][Three-body interaction. $R_{1}=1.6$]{\label{3bR2t1}\includegraphics[width=0.34\textwidth]{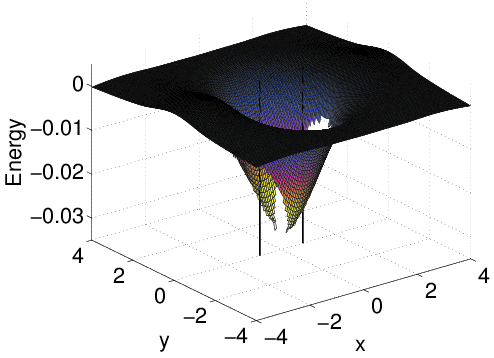}}
	\subfloat[][Three-body interaction. $R_{1}=2.0$]{\label{3bR3t1}\includegraphics[width=0.34\textwidth]{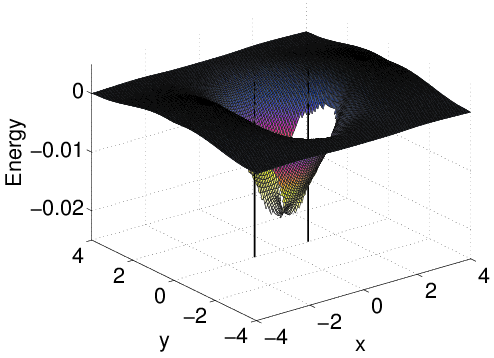}}
	\caption{Same type of data as in Fig. \ref{type1tb} but for a single component type-2 system with parameters $\alpha = -1$, $\beta=1$ and $q=1.5$.}
	\label{type2tb}
\end{figure*}

\begin{figure*}[!hbtp]
	\subfloat[][Total interaction. ]{\label{3bR1t1C}\includegraphics[width=0.34\textwidth]{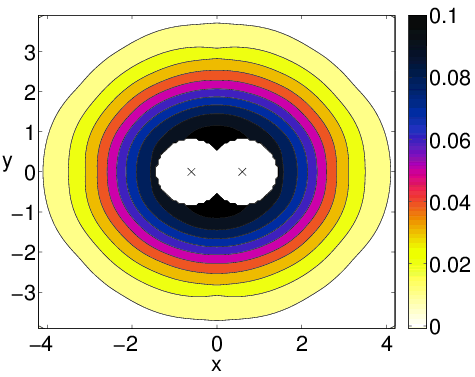}}
	\subfloat[][Total interaction. ]{\label{3bR1t1C}\includegraphics[width=0.34\textwidth]{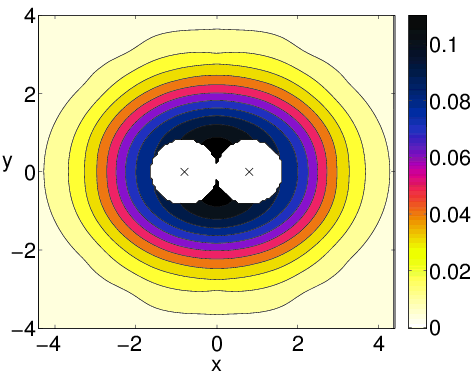}}
	\subfloat[][Total interaction. ]{\label{3bR1t1C}\includegraphics[width=0.34\textwidth]{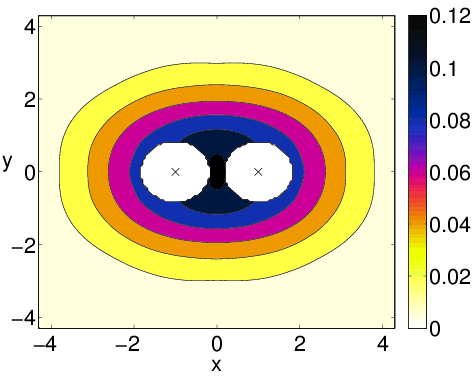}}\linebreak
	\subfloat[][Three-body interaction. $R_{1}=1.2$.]{\label{3bR1t1C}\includegraphics[width=0.34\textwidth]{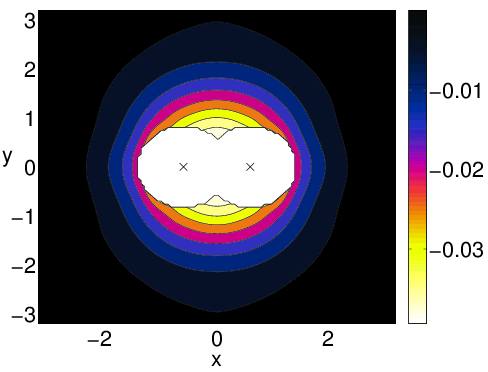}}
	\subfloat[][Three-body interaction. $R_{1}=1.6$.]{\label{3bR2t1C}\includegraphics[width=0.34\textwidth]{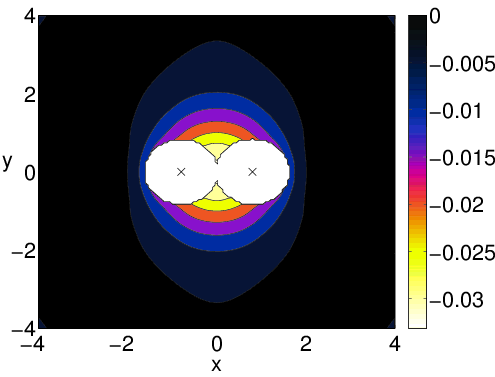}}
	\subfloat[][Three-body interaction. $R_{1}=2.0$.]{\label{3bR3t1C}\includegraphics[width=0.34\textwidth]{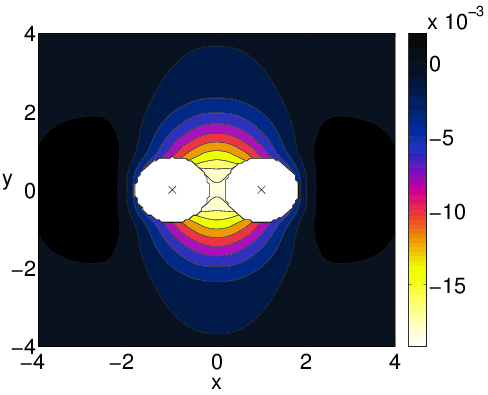}}
	\caption{Contour plots of the data from Fig. \ref{type2tb}.}
	\label{type2tbC}
\end{figure*}

Fig. \ref{type2tb} shows the total interaction, sum of pairwise interaction and three-body interaction energy of a single-component type-2 system with a first vortex pair placed at a distance $R_1$ from each other. The interaction energy is shown as a function of the position of the third vortex. The parameters used here are $\alpha = -1$, $\beta=1$ and $q=1.5$. Fig. \ref{type2tbC} shows the same data for total interaction and three-body interaction as that in Fig. \ref{type2tb} as contour plots.

In Fig. \ref{type2tb} it is seen that all the type-2 systems studied exhibit repulsive pairwise and total interactions consistent with known analytical and experimental results. The three-body interaction on the other hand is at all distances attractive causing a reduction in the total interaction energy compared to the sum of pairwise interaction. This is consistent  with three-body forces in type-2 two-component models \cite{PhysRevB.84.134515},  however, here a much wider set of configurations is studied.

\subsection{Two-Component Type-1.5 Three-Body Interaction}\label{sec:type15}

\begin{figure*}[!hbtp]
	\subfloat[][Total interaction.]{\label{3vR1}\includegraphics[width=0.34\textwidth]{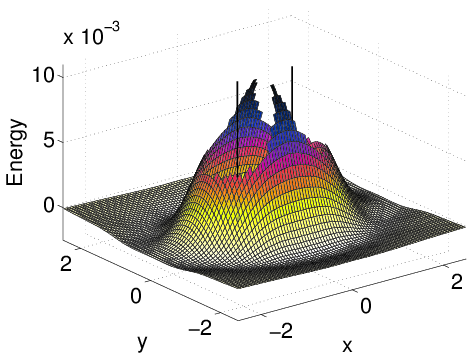}}
	\subfloat[][Total interaction.]{\label{3vR2}\includegraphics[width=0.34\textwidth]{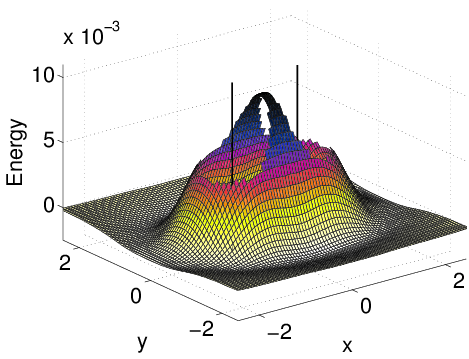}}
	\subfloat[][Total interaction.]{\label{3vR2}\includegraphics[width=0.34\textwidth]{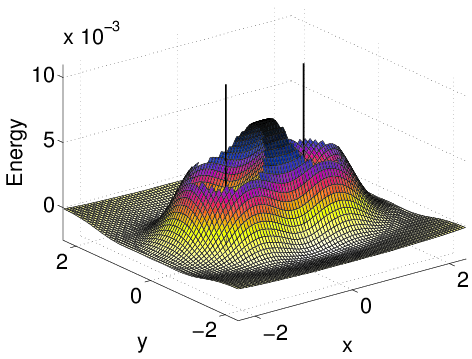}}\linebreak
	\subfloat[][Sum of pairwise interactions.]{\label{3vR1}\includegraphics[width=0.34\textwidth]{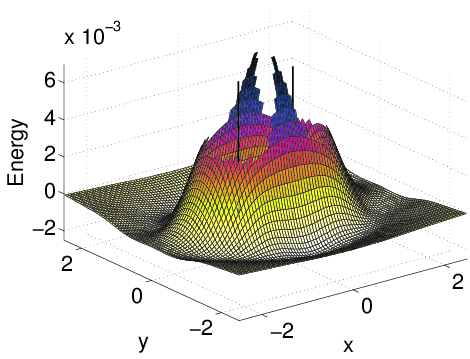}}
	\subfloat[][Sum of pairwise interactions.]{\label{3vR2}\includegraphics[width=0.34\textwidth]{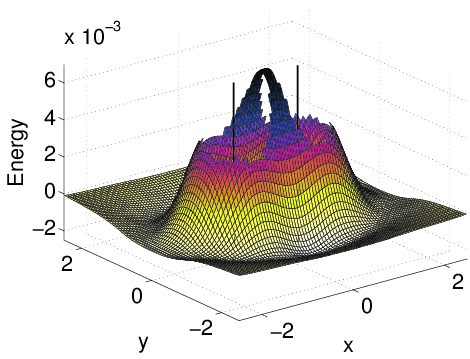}}
	\subfloat[][Sum of pairwise interactions.]{\label{3vR2}\includegraphics[width=0.34\textwidth]{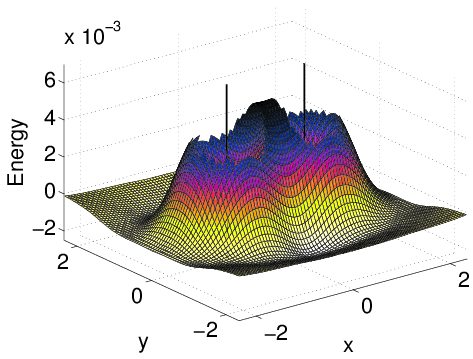}}\linebreak
	\subfloat[][Three-body interaction. $R_{1}=1.2$]{\label{3biR1typ15}\includegraphics[width=0.34\textwidth]{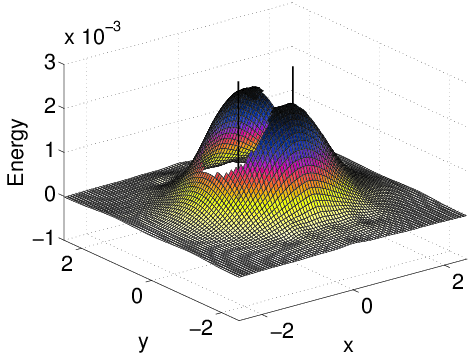}}
	\subfloat[][Three-body interaction. $R_{1}=1.4$]{\label{3vR2}\includegraphics[width=0.34\textwidth]{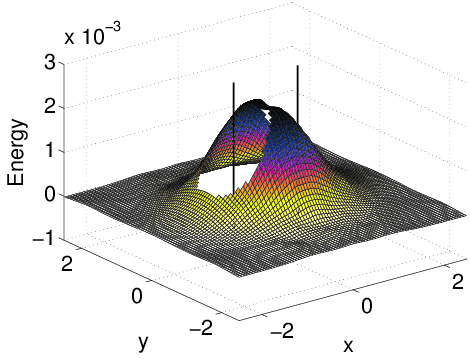}}
	\subfloat[][Three-body interaction. $R_{1}=1.6$]{\label{3vR2}\includegraphics[width=0.34\textwidth]{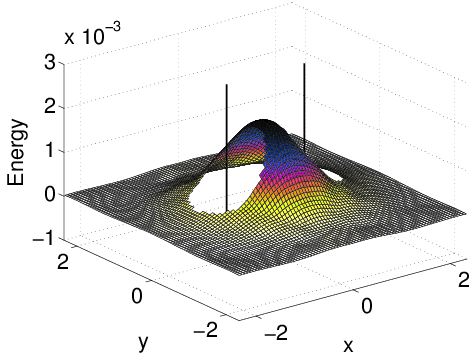}}
	\caption{Same type of data as in Fig. \ref{type1tb} and Fig. \ref{type2tb} but for a two-component type-1.5 system with parameters $\alpha_1 = -1.0$, $\beta_1 = 1.0$, $\alpha_2 = 3.0$, $\beta_2 = 0.5$, $q = 1.5$ and $\eta = 7.0$.}
	\label{type15tb}
\end{figure*}

\begin{figure*}[!hbtp]
	\subfloat[][Total interaction. ]{\label{3bR1t1C}\includegraphics[width=0.34\textwidth]{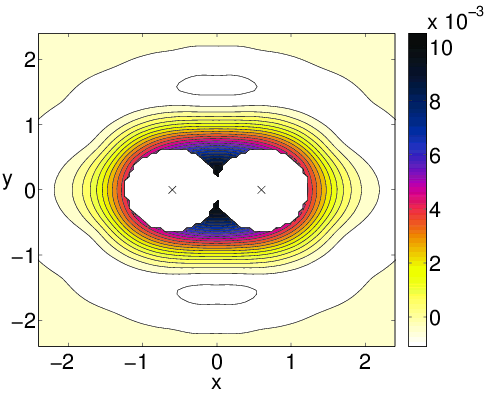}}
	\subfloat[][Total interaction. ]{\label{3bR1t1C}\includegraphics[width=0.34\textwidth]{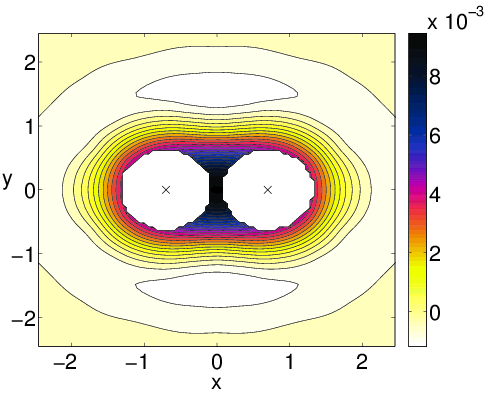}}
	\subfloat[][Total interaction. ]{\label{3bR1t1C}\includegraphics[width=0.34\textwidth]{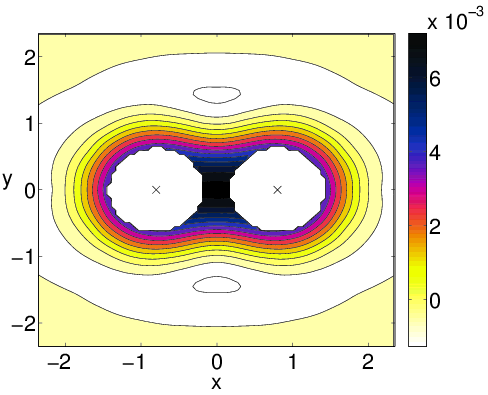}}\linebreak
	\subfloat[][Three-body interaction. $R_{1}=1.2$.]{\label{3bR1t2C}\includegraphics[width=0.34\textwidth]{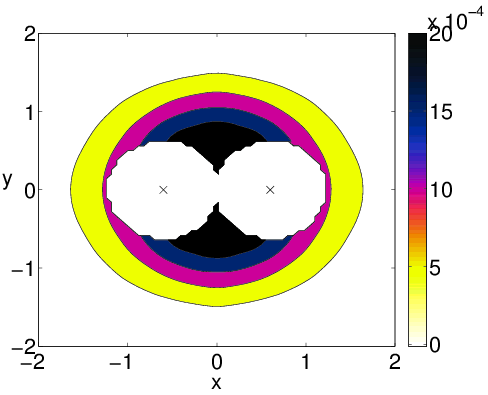}}
	\subfloat[][Three-body interaction. $R_{1}=1.4$.]{\label{3bR2t2C}\includegraphics[width=0.34\textwidth]{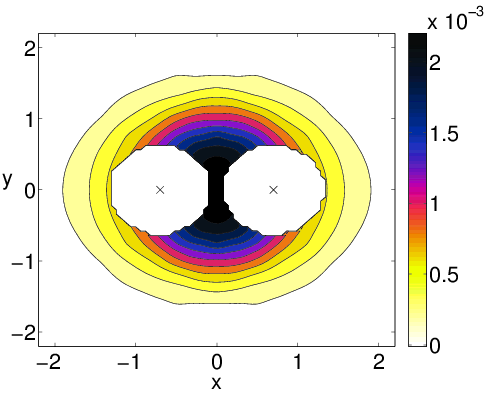}}
	\subfloat[][Three-body interaction. $R_{1}=1.6$.]{\label{3bR3t2C}\includegraphics[width=0.34\textwidth]{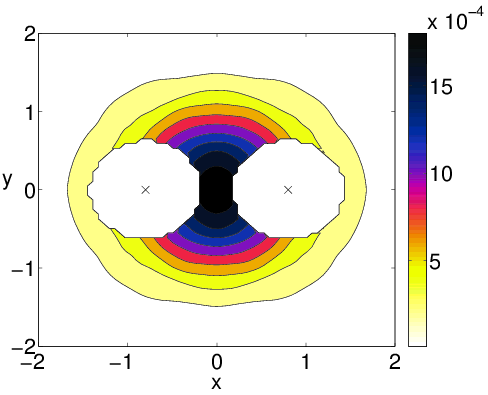}}
	\caption{Contour plots of the data from Fig. \ref{type15tb}.}
	\label{type15tbC}
\end{figure*}

Multibody forces in the two-component case should in general not be expect to be the same as in the single-component case.
This is because in the two-component case a vortex is a composite object made of two fractional vortices. For a review see \cite{Babaev20122}.
Thus the multibody forces which arise due to mutual deformation of vortices can include 
splitting of a composite vortex into constituent fractional vortices. The fractional vortices
have a number of exotic properties such as flux delocation and inversion \cite{PhysRevLett.103.237002} which makes it an especially complicated nonlinear problem.
Fig. \ref{type15tb} shows the total interaction, sum of pairwise interactions and three-body interaction energy of a two-component type-1.5 system with a first vortex pair placed at a distance $R_1$ from each other. The interaction energy is shown as a function of the position of the third vortex. The parameters used are $\alpha_1 = -1.0$, $\beta_1 = 1.0$, $\alpha_2 = 3.0$, $\beta_2 = 0.5$, $q = 1.5$ and $\eta = 7.0$. Fig. \ref{type15tbC} shows the same data for total interaction and three-body interaction as that in Fig. \ref{type15tb} but as contour plots.

All of the type-1.5 systems studied here show a pairwise and total interaction which is short-range repulsive and long-range attractive while the three-body interaction is at all distances repulsive. These observations are consistent with results in \cite{PhysRevB.84.134515} but a wider range of vortex configurations have been studied here, giving a complete picture of the three-body interaction. In Fig. \ref{type15tb} it can be seen how the maximum magnitude of the three-body interaction decreases with increasing $R_1$.

\subsection{Four-Body Interactions}\label{sec:4bint}

It should not in general be possible to describe the interaction between more than three vortices as neither the sum of pairwise interactions nor the sum of pairwise and three-body interactions. It is therefore of interest to investigate also the interactions in systems of more vortices and hence a four-body interaction is defined as the difference between the total interaction and the sum of pairwise and three-body interactions in a system of four vortices. Such interactions are studied here for the same single-component type-1 and type-2 and two-component type-1.5 systems as studied in the previous sections. As a system of four vortices has more degrees of freedom than three vortices, the case studied is limited to that of a square configuration. 

\begin{figure*}[!hbtp]
	\subfloat[][Type-1.]{\label{type1fourbb}\includegraphics[width=0.34\textwidth]{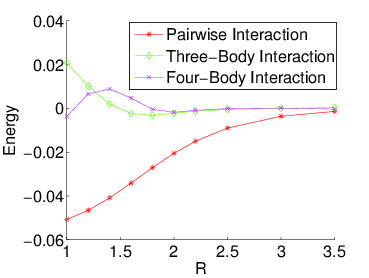}}
	\subfloat[][Type-2.]{\label{type2fourbb}\includegraphics[width=0.34\textwidth]{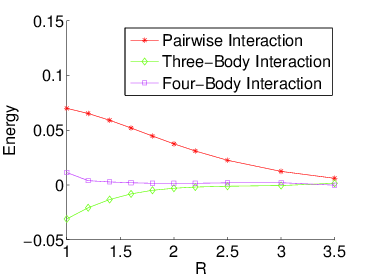}}
	\subfloat[][Type-1.5.]{\label{type15fourba}\includegraphics[width=0.34\textwidth]{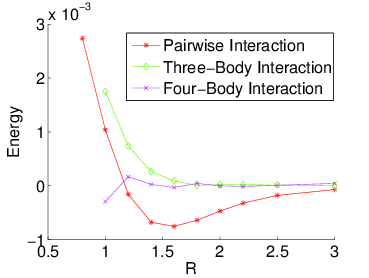}}\linebreak
	\subfloat[][Type-1.]{\label{type1fourbb}\includegraphics[width=0.34\textwidth]{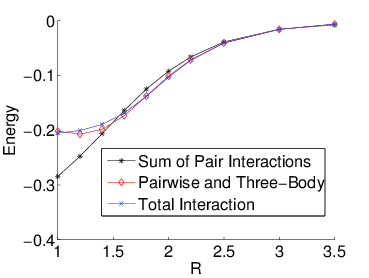}}
	\subfloat[][Type-2.]{\label{type2fourbb}\includegraphics[width=0.34\textwidth]{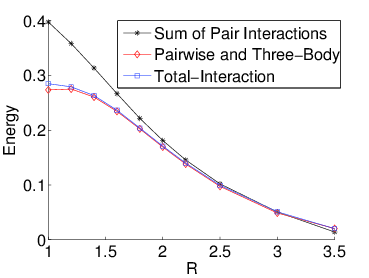}}
	\subfloat[][Type-1.5.]{\label{type15fourba}\includegraphics[width=0.34\textwidth]{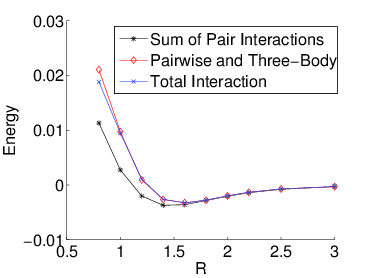}}
	\caption{Interaction energies for four vortices in a square with side $R$. Total interaction, sum of pairwise interactions and sum of pairwise and three-body interactions. (a)-(b) show single-component type-1 and type-2 systems with parameters $\alpha = -1$, $\beta=1$ and $q=2.5$ and $\alpha = -1$, $\beta=1$ and $q=1.5$ respectively. (c) is a type-1.5 system with parameters $\alpha_1 = -1.0$, $\beta_1 = 1.0$, $\alpha_2 = 3.0$, $\beta_2 = 0.5$, $q = 1.5$ and $\eta = 7.0$.}
\label{fourb}
\end{figure*}

Fig. \ref{fourb} shows interaction energies of a configuration of four vortices in a square with side length $R$. The top row shows interaction energy for a single pair with distance $R$, a single triangle with sides $R$, $R$ and $\sqrt{2}R$ as well as the four-body interaction of the square. The bottom row shows the total interaction of the square, the sum of pairwise interactions as well as the sum of pairwise and three-body interactions. Left column is the same single-component type-1 system previously studied with parameters $\alpha = -1$, $\beta=1$ and $q=2.5$. Middle column is a single-component type-2 system with parameters $\alpha = -1$, $\beta=1$ and $q=1.5$. Right column shows the two-component type-1.5 system with parameters $\alpha_1 = -1.0$, $\beta_1 = 1.0$, $\alpha_2 = 3.0$, $\beta_2 = 0.5$, $q = 1.5$ and $\eta = 7.0$. 

The most important conclusion to be drawn from Fig. \ref{fourb} is that, in the studied regimes, the four-body interaction is smaller than pairwise and three-body interactions, except possibly in the type-1 case where in some cases it is of similar size as the three-body interaction. In the bottom row of Fig. \ref{fourb} it is seen however that in all cases the approximation of the total interaction by the sum of pairwise and three-body interactions is significantly better compared to that of only the sum of pairwise interactions, especially at short distances. Adding also the four-body interaction does not make as big difference as the three-body interaction. This would imply that the three-body interaction is of greater importance than four-body interactions in the systems studied here. Even in the type-1 case where the four-body interaction seems stronger, it is not of as great importance since in a system of four vortices there is only one contribution of four-body interaction but four contributions of three-body interaction.

\section{Conclusions}\label{sec:concl}

Non-pairwise interactions between vortices in single component type-1 and type-2 as well as two-component type-1.5 superconductors have been studied numerically in Ginzburg-Landau theory. In the studied parameter sets the results show that there is a three-body interaction which is short-range repulsive but weakly long-range attractive in the type-1 case, attractive in the type-2 case and repulsive in the type-1.5 case. In the critical case, $\kappa=\frac{1}{\sqrt{2}}$, it is confirmed that just as there is no pairwise interaction, there are also no three-body interactions (within the accuracy of our numerical scheme). Results presented for systems of four vortices imply that the three-body interactions are of greater importance than four-body interactions so that inclusion of a three-body interaction might be an improvement to the approximation of the total interaction by the sum of pairwise interactions. The results are important for giant vortices in type-1 superconductors \cite{nphys888}, pinned clusters of vortices in type-2 superconductors and vortex cluster states in type-1.5 superconductors  \cite{Babaev20122,PhysRevB.84.134515}

I thank Johan Carlstr\"om for providing the code for computations and for discussions, Egor Babaev for suggesting the problem and discussions, and Physics Department of University of Massachusetts, Amherst, where a part of this work was done, for hospitality. This work was supported by Swedish Research Council. The computations were performed on resources provided by the Swedish National Infrastructure for Computing (SNIC) at National
Supercomputer Center at Link\"oping, Sweden.

\bibliography{literature}{}
\bibliographystyle{ieeetr}

\end{document}